\documentclass[pra,superscriptaddress,twocolumn]{revtex4}
\usepackage{dcolumn}
\usepackage{amssymb,amsmath}
\usepackage{graphicx}
\usepackage{hhline}
\usepackage{textcomp}
\usepackage{wasysym}
\usepackage{upgreek}
\usepackage{gensymb}
\usepackage{amssymb}
\usepackage{enumitem}
\setlist{noitemsep}

\newcommand{\appropto}{\mathrel{\vcenter{
  \offinterlineskip\halign{\hfil$##$\cr
    \propto\cr\noalign{\kern2pt}\sim\cr\noalign{\kern-2pt}}}}}

\begin{document}

\raggedbottom

\title{A Velocity Map Imaging apparatus optimized for high-resolution crossed molecular beam experiments}

\author{Vikram Plomp}
\affiliation{Radboud University Nijmegen, Institute for Molecules and Materials, Heijendaalseweg 135, 6525 AJ Nijmegen, the Netherlands}

\author{Zhi Gao}
\affiliation{Radboud University Nijmegen, Institute for Molecules and Materials, Heijendaalseweg 135, 6525 AJ Nijmegen, the Netherlands}

\author{Sebastiaan Y. T. van de Meerakker}
\affiliation{Radboud University Nijmegen, Institute for Molecules and Materials, Heijendaalseweg 135, 6525 AJ Nijmegen, the Netherlands}

\date{\today}

\begin{abstract}
We present the design of a Velocity Map Imaging apparatus tailored to the demands of high-resolution crossed molecular beam experiments employing Stark or Zeeman decelerators. The key requirements for these experiments consist of the combination of a high relative velocity resolution for large ionization volumes and a broad range of relatively low lab-frame velocities. The SIMION software package was employed to systematically optimize the electrode geometries and electrical configuration. The final design consists of a stack of 16 tubular electrodes, electrically connected with resistors, which is divided into three electric field regions. The resulting apparatus allows for an inherent velocity blurring of less than 1.1 m/s for NO$^+$ ions originating from a 3x3x3 mm ionization volume, which is negligible in a typical crossed beam experiment. The design was recently employed in several state of the art crossed-beam experiments, allowing the observation of fine details in the velocity distributions of the scattered molecules.
\end{abstract}

\maketitle
\section{Introduction}
Over the past 30 years, the technique of ion imaging has been acknowledged in a diverse selection of research fields, and various implementations have allowed for a multitude of applications of this powerful technique \cite{Ashfold2006}. The method was first reported by Chandler and Houston in 1987, where a relatively simple electrode configuration consisting of a repeller plate and two extractor grids was used \cite{Chandler1987}. By allowing the Newton sphere to expand during the flight towards a 2D-position-sensitive detector it was possible to study the velocity distribution of photodissociation fragments of CH$_3$I. However, the spread in initial ionization position severely limited the velocity resolution. A major improvement to the ion imaging technique was introduced by Eppink and Parker in 1997, in the form of the Velocity Map Imaging (VMI) detector employing electrostatic lenses \cite{Eppink1997}. The replacement of the electrode grids with apertures breaks the homogeneity of the electric fields at the transition between the electric field regions. This allows for focussing of particles with the same initial velocity vector \cite{Liebl2008}, but originating from different positions in the ionization region, on a single spot of the 2D-position-sensitive detector.

The Eppink-Parker VMI-detector design is still employed in its original form today. However, various modifications have been reported to either improve the performance of the technique or extend its use to different ranges of application. Most notably, the concept of image slicing was introduced, which was first explored by Gebhardt et al. \cite{Gebhardt2001}. Here, only a small part of the Newton sphere in the axial direction (2D-slice) is selected for detection at a time. Recording the 2D-velocity distribution of different slices then allows for direct measurements on the full 3D-velocity distribution of the products \cite{Ashfold2006}. Image slicing also greatly reduces the negative effects of Newton sphere crushing on the image resolution. Using this technique, experimental resolution down to 0.19\% (FHWM) has been reported in photodissociation experiments with a high kinetic energy release \cite{Lipciuc2006}.

In the field of crossed molecular beam experiments two frequently used VMI ion optics designs, on which many others are based, are those reported by Townsend et al. \cite{Townsend2003} and Lin et al. \cite{Lin2003}. The first employs three electric field regions, instead of the traditional two. Thus, the electric field strength in the ionization region could be reduced, allowing image slicing of photofragment velocity distributions while maintaining proper velocity mapping conditions. The second design uses only two electric field regions, but consists of a total of 29 electrode plates. The larger axial extent of the ion optics reduces the electric field strength in the ionization region, again enabling image slicing, as well as providing a softer focus which improves the velocity resolution.

In a crossed beam imaging experiment the inherent resolution is limited by the velocity spreads of the molecular beams. Recently, the employment of molecular decelerators in crossed beam experiments has enhanced the possibilities to investigate collisions between neutral species \cite{Onvlee2014}. The narrow velocity and angular spreads of Stark or Zeeman decelerated beams result in scattering images with unprecedented resolution, that can be exploited to resolve structure in the scattering images that would have been washed out using conventional molecular beams \cite{Zastrow2014,Gao2018pccp,Plomp2020,Jongh2020}. With this high inherent resolution, the experiments become more susceptible to imperfections in the velocity mapping. The sensitivity and resolution of these experiments thus not only depend on the control over the particles before the collision, but also on the quality of the velocity mapping apparatus.

Although the various optimized VMI-detector designs have improved the imaging resolution significantly, none of them were originally developed for the conditions encountered in controlled, high-resolution collision experiments employing decelerators. These experiments pose specific requirements on the mapping properties that are substantially different from those in e.g. photodissociation experiments. In this work, we present a VMI-detector design specifically optimized for crossed molecular beam experiments employing decelerators, where a special emphasis is required on high-resolution imaging for large detection volumes and a broad range of collision energies.

\section{Requirements}
\begin{figure}[!htb]
   \centering
    \resizebox{0.8\linewidth}{!}
    {\includegraphics{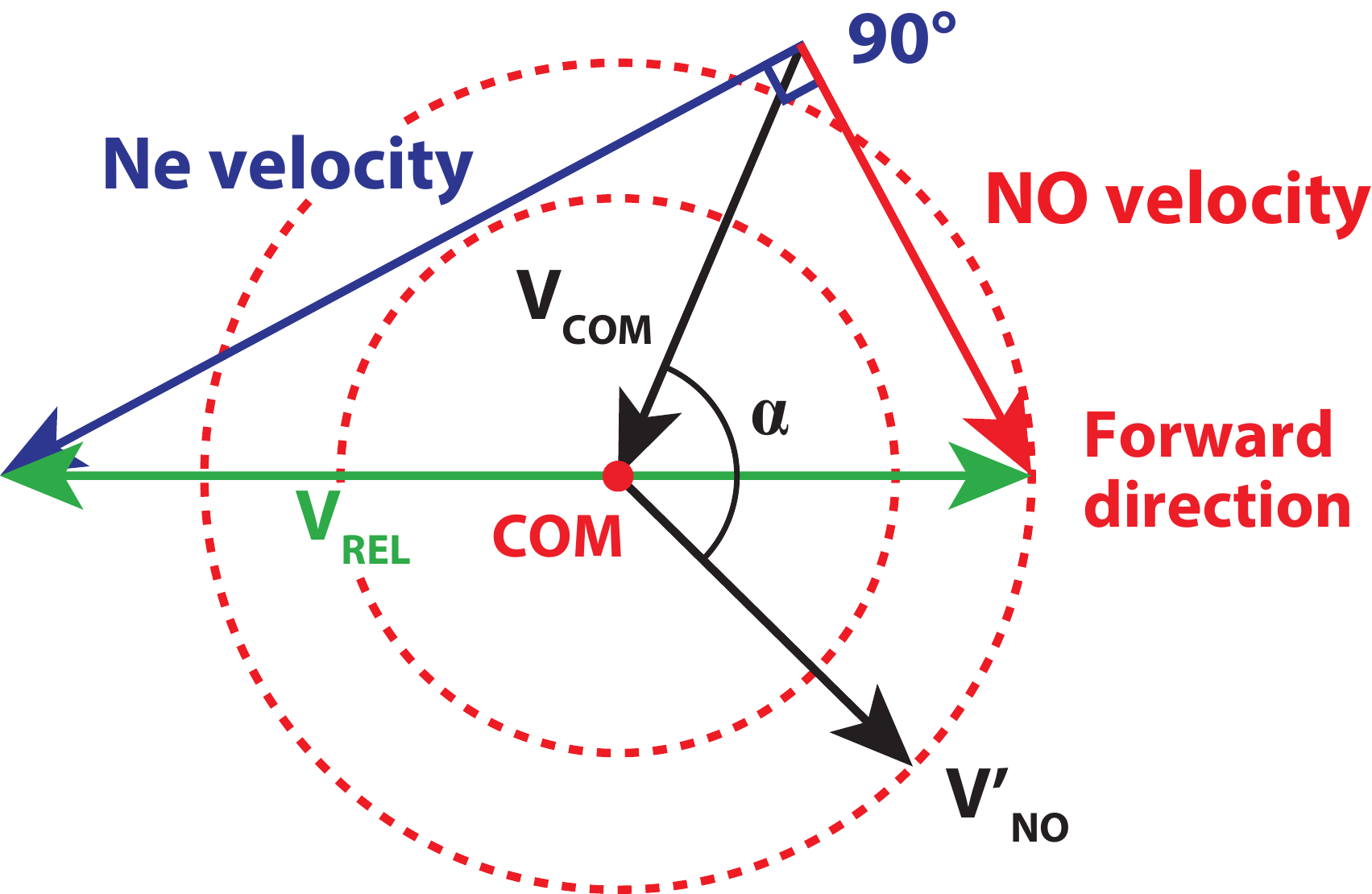}}
    \caption{Schematic 2D velocity diagram illustrating how the (collisionally excited) NO molecule velocity vectors are projected on Newton circles around the center of mass (COM) after scattering with Ne at an angle of $90\degree$. For a typical NO velocity of 430 m/s and Ne velocity of 800 m/s, corresponding to a collision energy of $\approx 410$ cm$^{-1}$, the COM-velocity amounts to $|V_{COM}|\approx411$ m/s, and the NO velocity in the COM-frame after elastic scattering is $|V'_{NO}|\approx363$ m/s. The angle between $V_{COM}$ and $V'_{NO}$ is $\alpha=67.1\degree$ for the forward scattering direction.}
    \label{fig:NewtDiag}
\end{figure}

We consider crossed beam experiments where at least one of the beams contains polar molecules, like NO, that can be manipulated using a decelerator. To illustrate the kinematics of such an experiment, we consider the collision of NO with Ne under typical conditions as an example system (see Fig. \ref{fig:NewtDiag} for the velocity diagram detailing the pre-  and post-collision conditions). The velocity of the scattered NO molecules is projected on concentric Newton spheres around the center off mass (COM) velocity ($V_{COM}$). Cylindrical symmetry of the experiment around the relative velocity axis ($V_{REL}$), however, ensures that the center slices of these spheres, the Newton circles, contain complete information on the scattering process \cite{Brouard2014}. The scattering distribution in both the radial and angular directions along these circles is governed by the interaction between the molecules during the collision. The narrow velocity spreads afforded by molecular decelerators allow the resolving of fine details in these scattering distributions \cite{Onvlee2014}. However, the Newton sphere radius in these experiments is generally very small. For the example NO-Ne collision system the 363 m/s radius of the Newton sphere corresponds to a kinetic energy of just 20.5 meV. This is significantly smaller than the typical $\approx1$ eV kinetic energy release in photodissociation experiments \cite{Townsend2003,Lin2003}, and provides the additional challenge to accurately map the velocities of these small Newton spheres.

The three main requirements of a VMI-apparatus for use in crossed molecular beam experiments employing decelerators, arise from the properties of the decelerators themselves:

\begin{enumerate}
\item Very small (relative) velocity blurring caused by the VMI ion optics, even at low ion kinetic energies.
\end{enumerate}

This constitutes the main requirement, as it is necessary to complement the high velocity resolution and broad range of collision energies afforded by the decelerator.

\begin{enumerate}[resume]
\item Accurate mapping of a large ionization volume.
\end{enumerate}

Although the particle density in phase-space is conserved throughout the deceleration process, the number of molecules that exit the decelerator is but a fraction of those in the initial molecular beam \cite{Meerakker2012}. In combination with the relatively small cross-sections of most neutral-neutral interactions, low ion yields can thus be expected. An enlarged ionization volume increases the amount of scattered molecules that can potentially be detected. Furthermore, the larger volume helps to reduce velocity blurring due to space charge effects under experimental conditions that do allow for higher ion yields \cite{Ashfold2006}. A large ionization volume, however, requires a better focussing of the ion optics, and is therefore in direct conflict with the requirement of small velocity blurring. The molecular beam that exits a Stark or Zeeman decelerator has a typical diameter of 2-3 mm \cite{Meerakker2012,Cremers2019}, which provides an upper limit to the practical size of the interaction volume.

\begin{enumerate}[resume]
\item Low electric field strength in the ionization region.
\end{enumerate}

The use of decelerators ensures high quantum state purity of the molecular beams, which allows for near background free measurement of state-to-state scattering processes. However, for several relevant species this state purity is compromised by the electric fields present in the detection region. While a Stark decelerator, for example, only transmits NO molecules with \emph{f} parity, an electric field mixes the close-lying Lambda-doublet states with \emph{e} and \emph{f} parity. Although the mixing probabilities are generally very small, their effects can be observed in experimental scattering images. For the conditions in a typical VMI-detector, the parity mixing probability depends quadratically on the electric field strength. Thus, reduced electric field strengths in the ionization region strongly suppress the possible influence of quantum state mixing. In this regard, the method of delayed extraction seems ideal, where pulsed voltages are applied to the VMI-apparatus after ionization in field free conditions. However, this method generally leads to increased velocity blurring due to the practical limitations of high-voltage switching, leaving the application of DC voltages the preferred method to obtain minimal blurring \cite{Chestakov2004,Ashfold2006}.\\

Further requirements to the mapping properties constitute of:
\begin{itemize}
\item Simultaneous mapping of a large range of ion velocities with high resolution.
\item Minimal dependence of the velocity resolution on the azimuthal angle ($\alpha$) of the particle on the Newton sphere.
\item Linear velocity mapping behaviour for a large range of ion velocities.
\item Sufficient total ion acceleration to ensure high detection efficiency of the employed Micro-Channel Plate (MCP) detectors.
\end{itemize}

Additional constraints to the dimensions of the VMI-apparatus are imposed for practical reasons:
\begin{itemize}
\item The ion optics should fit within a cylinder with 80 mm diameter.
\item The ion optics should not exceed a height of 300 mm above the center of the detection region.
\end{itemize}

It should be noted that the requirement of a low electric field strength in the ionization region is closely related to the capability of direct image slicing, as it determines the total elongation of the Newton sphere. However, accurate image slicing also requires focussing in arrival-time at the detector for particles starting at different positions in the ionization volume, to minimize the effective slice thickness. For the small Newton spheres encountered in low-energy collision experiments, this time-focussing requirement is especially strict, and fundamentally conflicts with low blurring for a large ionization volume \cite{Ryazanov2013}. While image slicing is greatly beneficial, it was found subordinate to the aforementioned requirements for the case of low energy crossed beam experiments employing a decelerator. The need for direct image slicing is also reduced by a continuously improving toolbox of general 3D-image reconstruction techniques \cite{Ashfold2006}, with recent additions of e.g. the FINA software package \cite{Thompson2017}.

\section{Electrostatic lens aberrations}
As the main requirements to the VMI-setup entail a small intrinsic velocity blurring for a large ionization volume, it is insightful to discuss the origins of this velocity blurring. The radial focussing of particles with the same initial velocity vector is governed by the electrostatic lenses at the borders of the (axial) electric field regions in a VMI-apparatus. These electrostatic lenses mainly suffer from two types of imperfections, namely chromatic aberrations and spherical (geometric) aberrations \cite{Liebl2008}.

Chromatic aberrations originate from the fact that the focal length of the lenses is dependent on the ion kinetic energy at the lens plane, similar to the dependence of the focal length of an optical lens on the wavelength of light. The extent of the aberrations depends on the relative difference in kinetic energy of the ions at the lens plane, which can arise from either different initial velocity or different ionization position (i.e. creation at a different electrostatic potential). These aberrations are generally largest for the first lens encountered, since the relative kinetic energy difference will be largest here.

Spherical aberrations are caused by a dependence of the focal length on the radial distance ($r$) from the lens axis, i.e. the radial field strength is not perfectly linear with respect to $r$. Since the Newton sphere expands when moving through the VMI-apparatus, the effect is generally largest for the last lens(es) encountered.

Chromatic aberrations can be reduced by decreasing the relative kinetic energy difference of the ions. Besides the use of a smaller ionization volume, this can effectively be achieved by moving away the first lens from the detection region (i.e. to either obtain a smaller field strength in the ionization region for similar potential at the lens plane, or obtain larger total ion acceleration before the first lens for similar field strengths). Also a soft (gradual) focus reduces the effect of the change in focal length of the first lens(es) on the image quality. These approaches are demonstrated in for example the designs of Townsend et al. \cite{Townsend2003} and Lin et al. \cite{Lin2003}.

Spherical aberrations can be reduced by decreasing the expansion of the Newton sphere (e.g. by using a larger overall field strength or shorter acceleration region), increasing the lens diameter, or using gradually focussing lenses with a larger axial extent of each individual lens in the ion optics system \cite{Liebl2008}. Also the use of a smaller ionization volume can help to reduce spherical aberrations, as this reduces the initial radius of the ion cloud.

It is apparent that the two types of aberrations can generally not be minimized simultaneously, as they for example have an opposite dependence on the length of the first field region. Thus, their combined minimum needs to be found. For the conditions encountered in low-energy crossed beam experiments, the chromatic aberrations tend to dominate as small Newton spheres and large ionization volumes are considered. Generally, only increasing detector size (uniform scaling), reducing the size of the ionization volume, and more gradual focusing, suppress both aberrations without having other negative effects on the mapping properties.

The electrode shape plays an important role in the mapping characteristics. Several efforts have been made to adjust the traditional flat annular electrodes in order to increase the mapping accuracy \cite{Ashfold2006}, for instance to improve the performance in crossed beam ion-molecule reaction studies \cite{Trippel2009}. Instead of the traditional electrode plates, also cylindrical or tube-shaped electrodes can be used. These have the advantage of reduced spherical aberrations due to an increased effective aperture for the same lens diameter, as well as a larger axial extent of the lens fields \cite{Liebl2008}. Another benefit is that for small inter-tube gaps, the external electric field penetration into the tubes is negligible, effectively shielding the VMI-region from outside influences. The use of tube-shaped electrodes was already suggested in the original work of Eppink and Parker \cite{Eppink1997}, and has been previously implemented \cite{OorschotThesis}. However, their use in VMI-applications is seldom reported, although a combination of planar-shaped electrodes with shielding tubes is implemented in multiple designs \cite{Trippel2009,Ghafur2009}.

The freedom in the design parameters, i.e. the electrode geometries and applied voltages, can be used to find a combined minimum of the aberrations and simultaneously satisfy the other mapping requirements. Adding degrees of freedom, like additional field regions \cite{Townsend2003} or magnifying lenses \cite{Offerhaus2001,Zhang2011,Ryazanov2013}, can allow for further control over the mapping properties. Due to the complexity of the system, the performance characteristics of a VMI-setup can in general not be readily calculated analytically. Instead, one can make use of numerical ion trajectory simulations as a method to predict and optimize the performance of a VMI-apparatus.

\section{Simulation routine}
The SIMION software package \cite{Dahl2000} (version 8.1) was utilized to simulate ion trajectories, and predict the performance of a VMI-setup. To achieve the best performance regarding our design goals, many possible VMI-setups were simulated and analyzed. Employing the user programming features of SIMION, a systematic approach was implemented to change and optimize both the geometrical and electrical configurations. The concepts to reduce lens aberrations, as well as previous ion optics designs, were used as an inspiration and to provide a starting point for the optimization process.

The global structure of the optimization routine consists of two main parts. The first part optimizes the voltage configuration of the VMI-setup for a single given geometrical configuration, which results in optimized performance for that geometry. To this extent the SIMION built-in implementation of a downhill Nelder-mead Simplex algorithm is used, in a similar fashion as in the work of Ryazanov et al. \cite{Ryazanov2013,RyazanovThesis}. It uses the applied electric potentials as input parameters, and finds a (local) minimum of an objective function that takes into account the three most important performance requirements, i.e. the velocity blurring for a given size ionization volume and the electric field strength in the detection area. The second part of the routine is used to optimize the geometry of the VMI-setup. It loops through a list of geometry variations, each time building the fully parametrized geometry file as well as the corresponding potential array. The best performance for each geometry is found through optimization of the electrode voltages, and stored for later analysis and further characterization. In this way, a reasonable part of the very high-dimensional parameter space can be explored without the use of advanced and computationally heavy optimization algorithms.

We used the example NO-Ne collision system (see Fig. \ref{fig:NewtDiag}) as a benchmark system in the optimization process. The requirement of a large detection volume is safeguarded by spawning the NO$^+$ ions at 125 positions uniformly distributed over a 3x3x3 mm cube for each velocity vector considered. The intrinsic blurring of the VMI-apparatus $\sigma(V'_{NO},V_{COM},\alpha)$ is characterized from the ion trajectories of all particles with the same initial velocity vector by using the maximum spreads in their hit positions on the detector in both 2D-coordinates ($D_{y},D_{z}$) and their average mapping radius from the center of the sphere ($R$) as:
\begin{equation}
\sigma(V'_{NO},V_{COM},\alpha)=\frac{\sqrt{D_{y}^2+D_{z}^2}}{R}
\label{blurringformula}
\end{equation}
which is proportional to the intrinsic relative velocity resolution ($D_V/V'_{NO}$) for assumed linear velocity mapping. Here, $D_V$ indicates the maximum spread in the obtained velocity vector after mapping, i.e. the intrinsic absolute velocity resolution. It is worth noting that, due to its definition, $\sigma$ increases for smaller Newton rings. Furthermore, by taking the maximum spreads in the hit positions and adding the two dimensions in quadrature, the effective blurring is systematically overestimated. The definition, however, does not take into account the size of the MCP pores and camera pixels, or effects of (partial) Newton sphere crushing, that could limit the practical mapping resolution. For each starting position of the ion, several initial velocity vectors on differently sized Newton spheres are probed around their common COM-velocity vector.

While the simulations are performed for NO$^+$ ions, most results can easily be generalized to other species. This, since two ions with the same starting position, velocity direction and kinetic energy per unit charge, will have identically shaped trajectories in electrostatic fields \cite{SIMIONmanual}. Only the time in which these trajectories are traversed will depend on the mass.

Cylindrical symmetry of the electrodes is assumed in the geometry optimization process to reduce the computation time. The surroundings of the ion optics and flight region are included in the geometry definition. The distance from the center of the ionization region to the MCP-detector, was fixed to the value of 892.5 mm in the existing crossed beam setup. The total voltage drop over the VMI-apparatus was kept at 2000 V in the simulations.

A substantial amount of variations to the VMI-detector design was investigated which included, among other things, the amount of electric field regions (independent applied potentials), the field region lengths, number of individual electrodes and electrode shapes. Careful optimization of the VMI-detector geometry was performed. However, the addition of degrees of freedom that allowed for only a marginally better performance, but result in a great increase in complexity of the design, was prevented.

\section{Optimized design}
\begin{figure}[!htb]
   \centering
    \resizebox{1.0\linewidth}{!}
    {\includegraphics{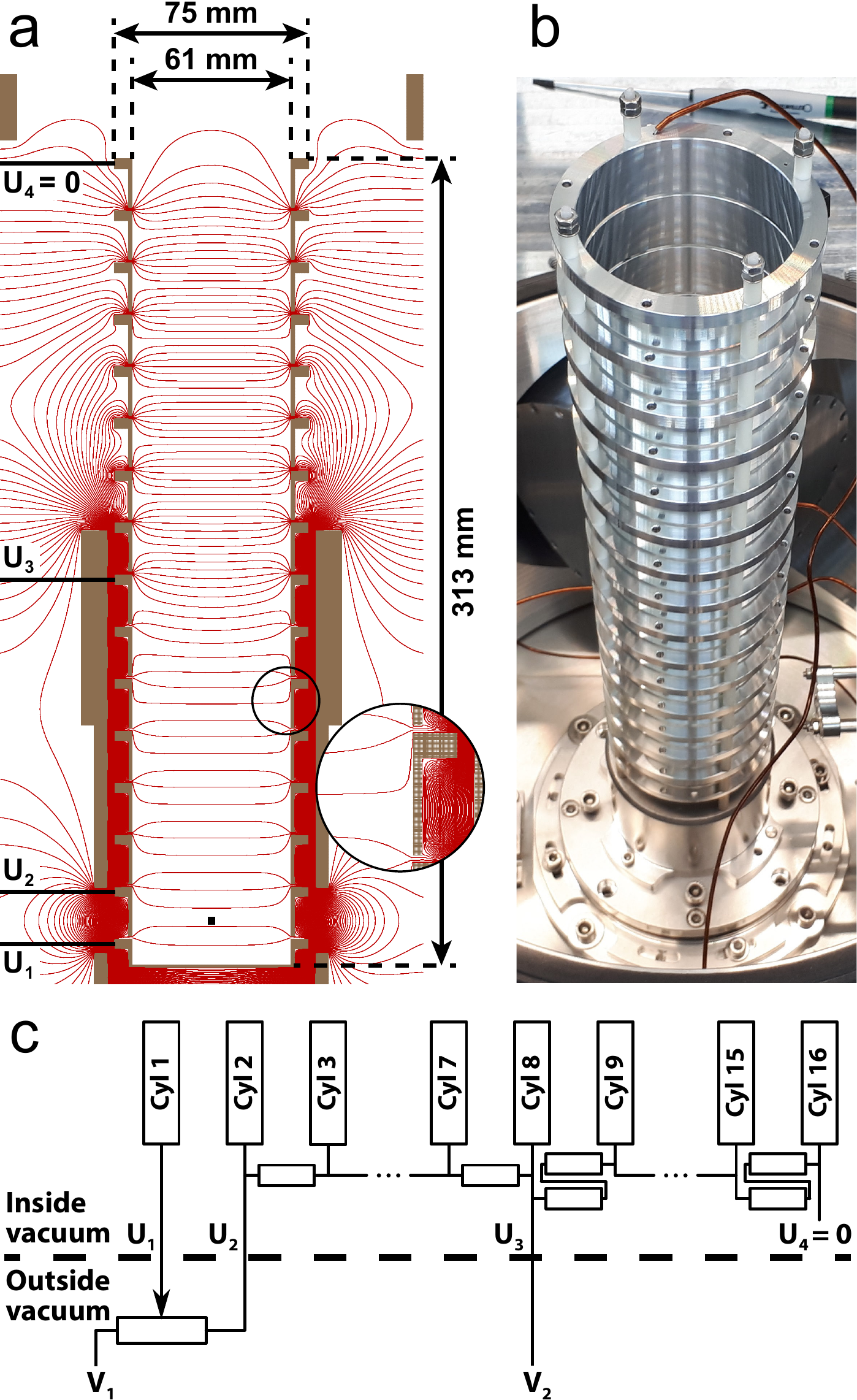}}
    \caption{(a) Cross-section of the optimized ion optics design and corresponding electric potential field as simulated with SIMION. The drawn equipotential lines have a 25 V spacing, with a total voltage drop of 2000 V over all cylinders. The center of the ionization volume is indicated by the black square. (b) Photograph of the assembled ion optics design. (c) Electrical scheme used to provide the correct potential to each cylinder (Cyl), employing two external power supplies (V$_1$, V$_2$), a 200 k$\Omega$ potentiometer and 100 k$\Omega$ resistors.}
    \label{fig:design}
\end{figure}

The optimized geometrical and electrical configuration, is shown in figure \ref{fig:design}. It consists of 16 cylinders with an inner diameter of 61 mm and outer diameter of 64 mm. The rims ($\diameter$75 mm, thickness 5 mm) are added for mounting only, and have very small influence on the mapping performance due to the rigorous shielding provided by the cylinders. Of the cylinders, 15 are identical with a length of 19 mm. Only the first cylinder is different, since it is closed on one side by the repeller plate. It has a total length of 13 mm with a 3-mm thick repeller plate. The surface of the repeller plate is positioned 17 mm away from the center of the ionization region. All cylinders have a 1-mm spacing in between, to ensure electrical isolation. The total distance from the center of the ionization region to the MCP-detector is set to 892.5 mm. The required beam access holes in the second cylinder ($\diameter$5 mm, every $45\degree$) are not shown. They break cylindrical symmetry, and their effect on the mapping accuracy will be separately addressed at the end of this section. Only four independent electric potentials are applied to the system, namely on the first cylinder/repeller (U$_1=2000$ V), the second cylinder (U$_2=1933.9$ V), the eight cylinder (U$_3=1472$ V) and the last cylinder (U$_4=0$ V). The other electrodes are connected via high precision 100 k$\Omega$ ($\pm0.01\%$) resistors, resulting in equal voltage steps within each region.

The resulting design is relatively simple and easy to build, but implements most of the discussed strategies to reduce lens aberrations. The most striking feature is the use of tubular electrodes instead of the traditional electrode plates with apertures or plates with shielding rims. They were found to provide the largest effective lens diameter and larger axial extent of the individual lenses, providing reduced mapping aberrations. Furthermore, it was found that while for electrode plates the spacing between the plates is an important parameter, the length of the individual cylinders has a much smaller influence on the mapping performance. This is expected to arise from the smoother change in the potential along the axis. Thus, longer cylinders could be used to reduce the amount of individual electrodes, greatly simplifying the design.

Three electric field regions were employed, as it was found that this provides a large increase in performance with respect to the traditional two field regions. However, the use of four electric field regions did not show a significant increase in mapping performance with respect to three field regions with optimized lengths, and would thus needlessly complicate the design and practical use of the apparatus.

The repeller plate was moved away from the ionization region to reduce the influence of the exact shape and positioning of the repeller plate on the mapping performance. Furthermore, separate control over the voltages on first and second cylinder allows for a slightly different field strength before and after the ionization region. This introduces a small, controllable, curvature of the electric field in the ionization region, which can help to improve focusing \cite{RyazanovThesis}. Additionally, this extra control over the field in the detection region may allow to correct for possible distortions introduced by the necessary access holes required for lasers and molecular beams. The transition between the second and third field region was moved to around cylinder eight, which allows for a low electric field strength in the ionization region and strongly reduces chromatic aberrations.

\begin{figure}[!htb]
   \centering
    \resizebox{1.0\linewidth}{!}
    {\includegraphics{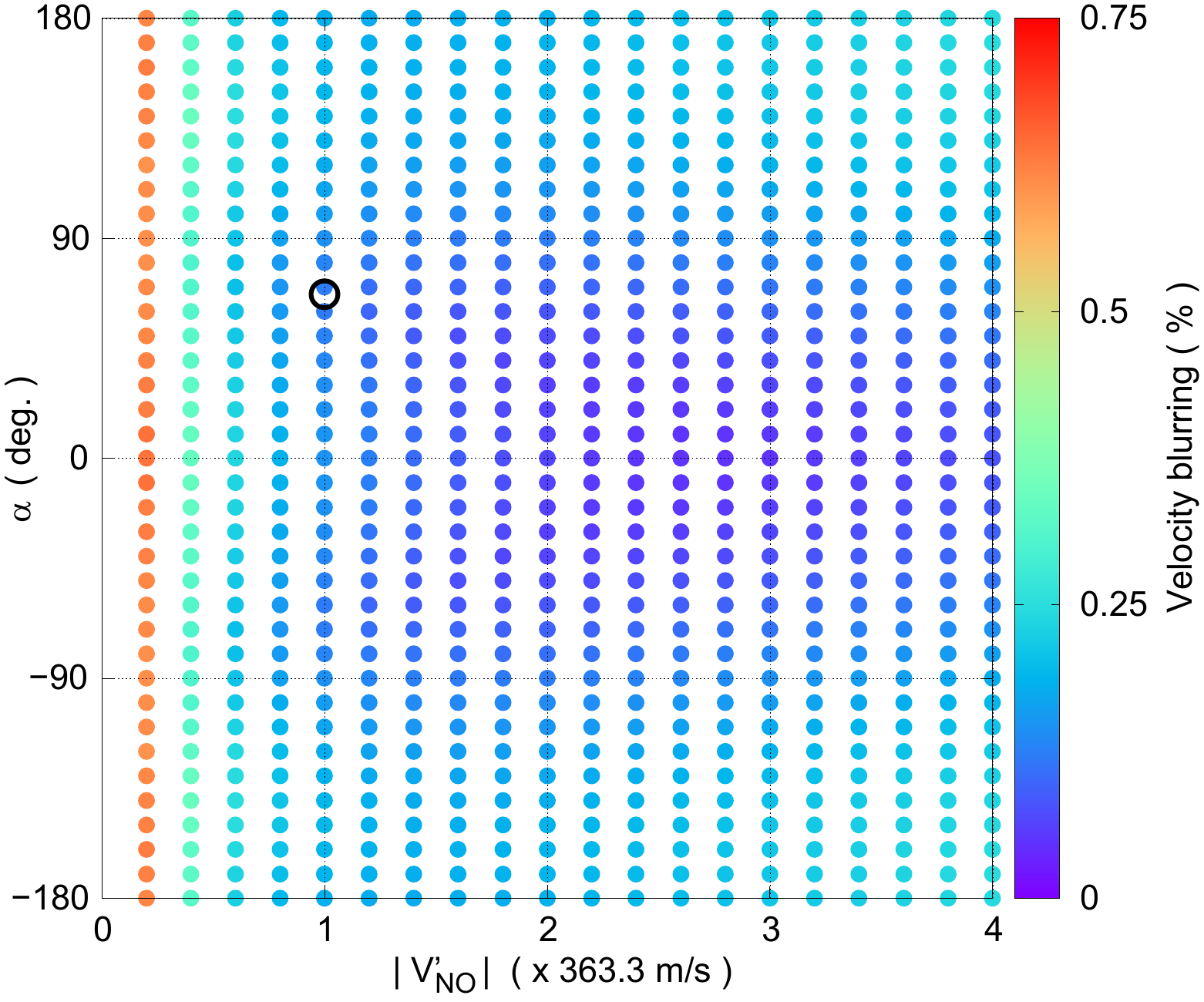}}
    \caption{Simulated velocity blurring ($\sigma$, see Eq. \ref{blurringformula}) of the optimized VMI-apparatus when using a 3x3x3 mm ionization volume, for varying Newton circle radius ($|V'_\text{NO}|$), and azimuthal angle ($\alpha$) with respect to a fixed COM-velocity ($|V_{COM}|=411.1$ m/s). The voltage settings were optimized on the forward scattering direction of the elastic NO-Ne reference collision system, indicated by the black circle ($|V'_\text{NO}|=363.3$ m/s, $\alpha=67.1\degree$, see Fig. \ref{fig:NewtDiag}).}
    \label{fig:blurring}
\end{figure}

The simulated blurring ($\sigma$, see Eq. \ref{blurringformula}) for the velocity mapping device is depicted in figure \ref{fig:blurring} for a variety of Newton circle radii ($V'_{NO}$) and azimuthal angles ($\alpha$) with respect to a fixed COM-velocity vector. Considering the forward scattering direction of the reference NO-Ne collision system, the simulated blurring for the 3x3x3 mm detection volume amounts to around only 3.14 $\upmu$m in both directions on the detector plane ($D_{y},D_{z}$), which is just below the 6 $\upmu$m center-to-center pore distance of the typically employed MCP-detector. Combined with an image radius of 3.806 mm this results in $\sigma=0.117 \%$ according to the definition of Eq. \ref{blurringformula}. Furthermore, the induced blurring is rather constant for all scattering angles, and remains very low even for the smallest investigated Newton sphere diameters. Thus, the new design was found to provide a significant improvement with respect to simulated benchmark designs with a more traditional lens arrangement.

It should be noted that the image radius ($R$) can be easily increased, with but a marginal increase in velocity blurring, by increasing the total flight length ($L$) or scaling down the applied voltages ($E$) as they are roughly related as $R\propto L\sqrt{\frac{m}{E}}\cdot |V'|$, where $m$ indicates the particle mass and $|V'|$ the velocity radius of the Newton sphere. Furthermore, the blurring is reduced for smaller ionization volumes. It was found to change similarly with an adjustment of the volume dimension along the axis of the VMI-apparatus, as with an equal adjustment in both the perpendicular dimensions. E.g. for a 1x1x1 mm volume the total blurring is reduced to $\sigma=0.017 \%$ for the forward scattering direction of the reference collision system. The mapping was furthermore found to be linear within 0.24\% with respect to the NO-Ne reference image radius, for all investigated Newton sphere diameters and scattering angles. This again provides a significant improvement with respect to simulated designs with a more traditional lens arrangement.

The electric field strength in the ionization region was found to be 35.10 V/cm for optimized voltage settings, and a voltage drop of 2000 V over all 16 cylinders. The obtained field strength should for example result in only very minor mixing of the NO ($X\,^2\Pi_{1/2}$) Lambda-doublet states, that is negligible in a typical experiment. The corresponding total Newton sphere elongation ($\tau$) for the reference scattering system is 64.4 ns, which is independent of the size of the ionization volume, and in principle allows for direct image slicing. As mentioned, however, the arrival-time focussing was not optimized for large ionization volumes, resulting in a fairly large time-blurring (maximum arrival-time spread $D_t$) of 47.3 ns for the 3x3x3 mm volume. Again, this blurring is heavily suppressed for smaller detection volumes, as it roughly scales with the dimension of the ionization volume along the axis of the VMI-apparatus. If the NO$^+$ ions originate e.g. from a 0.5x3x3 mm volume it amounts to around $D_t =$ 7.9 ns. Scaling down of the applied voltages ($E$) can be used to increase total Newton sphere elongation ($\tau$), as they are approximately related as $\tau\propto\frac{m|V'|}{E}$, where $m$ indicates the particle mass and $|V'|$ the velocity radius of the Newton sphere. The time-blurring for a given size ionization volume is almost independent of the velocity radius, and is loosely related to $D_t\appropto\sqrt{\frac{m}{E}}$. Lastly, the so called center slice distortion, given by the difference in mean arrival time for the centre slices of Newton spheres with different radii, was found to be negligible. Thus, direct image slicing can still be in reach under certain experimental conditions.

\begin{figure}[!htb]
   \centering
    \resizebox{1.0\linewidth}{!}
    {\includegraphics{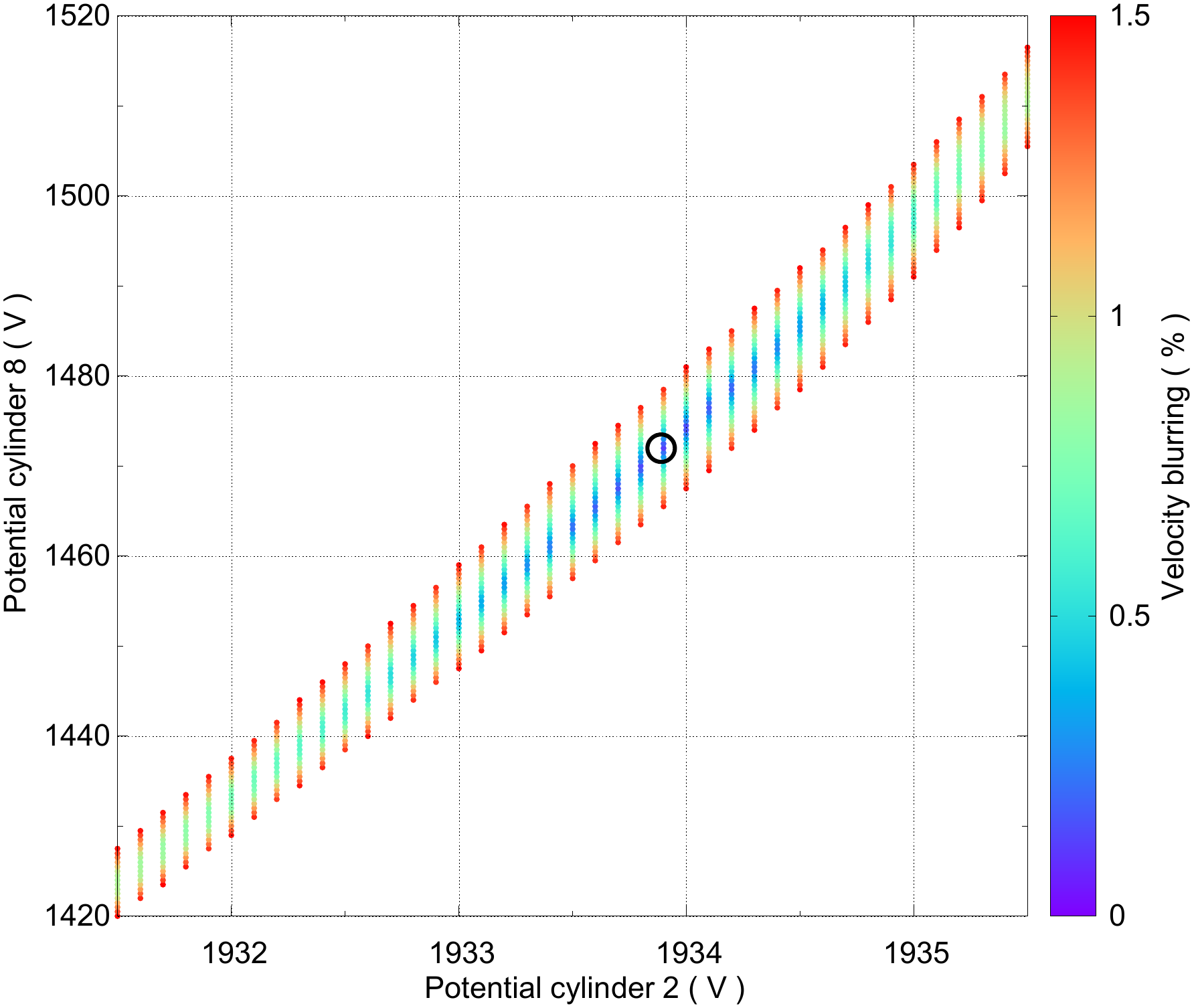}}
    \caption{Simulated velocity blurring ($\sigma$, see Eq. \ref{blurringformula}) of the optimized VMI-apparatus when using a 3x3x3 mm ionization volume, for varying voltages applied to the device. The potential applied to the first and last cylinder is fixed to 2000 V and 0 V, respectively. The blurring was determined for the forward scattering direction of the elastic NO-Ne reference collision system (see Fig. \ref{fig:NewtDiag}), for which the optimized voltage setting is indicated by the black circle.}
    \label{fig:voltages}
\end{figure}

The stability of the optimized design was investigated regarding deviations in both the geometry and applied voltages. These deviations are only expected to significantly affect the induced velocity blurring. Thus, to investigate the voltage stability the velocity blurring was mapped out for a range of voltages in the vicinity of the optimized settings. As the lens properties only depend on the ratio of the electric field strengths of succeeding field regions, only the potentials applied to the second and eight cylinder have to be investigated. The results are depicted in figure \ref{fig:voltages}.

It can be seen that a narrow voltage band exists where perfect mapping conditions are ensured. The accuracy of commercially available high voltage power supplies is sufficient to ensure proper settings for the potential on cylinder 8. For the potential on cylinder 2, however, the range of good voltage settings is even narrower. This could be expected as the first field region consists of only a single cylinder while it has the lowest electric field strength. This narrow voltage acceptance is especially challenging as most commercial power supplies have voltage output ripples and drifts of similar magnitude. For the power supplies employed here (SRS P350) these amount to maximum 0.1 V and 0.6 V (over 8 hours), respectively. The problem is easily solved, however, with the use of a potentiometer to set the voltage on the first cylinder and second cylinder with a single power supply (see Fig. \ref{fig:design}). In our implementation a 200 k$\Omega$ 10-turn precision potentiometer was used. This firstly allows for a much more precise setting of the voltage. Furthermore, it effectively fixes the ratio of the electric field strengths in the first and second region, thus conserving the focussing power of the first lens. Any voltage shift or ripple is spread out over the first two electric field regions, and can thus effectively be considered to accumulate on the set voltage of cylinder 8. This voltage has a much broader acceptance range, allowing stable operation with minimal velocity blurring.

It should be noted, that a voltage shift in principle not only affects the intrinsic velocity blurring ($\sigma$), but can also slightly change the center position and radius of the image on the detector plane. This can cause additional blurring that is not incorporated in $\sigma$. However, it was found from the simulations that these effects are significantly smaller than the change in the intrinsic velocity blurring ($\sigma$) for the large ionization volume and expected voltage shifts.

Regarding the stability with respect to deviations from the proposed geometry, two effects were investigated. Firstly, the VMI performance was simulated for a system where a combination of geometrical defects was deliberately added. It was found that, for reasonable errors in machining and assembly, only minor changes in the VMI performance could be expected. Errors in the inner diameter of the cylinders cause the largest changes in performance, which was accounted for during machining of the electrodes. Secondly, the influence of the required beam access holes in the second cylinder was investigated in 3D-simulations that take into account the breaking of the cylindrical symmetry. Beam access holes with a 5-mm diameter (every $45\degree$) allow sufficient room for access with both laser beams and molecular beams, while showing only marginal effects on the simulated mapping performance.

\section{Experimental}
A VMI-apparatus following the design described above was constructed out of aluminum, and implemented into an existing crossed beam scattering setup. This setup has been discussed in detail before \cite{Gao2018pccp}, and consists of a 2.6-m long Stark decelerator and a conventional beam at a 90$\degree$ crossing angle. A molecular beam of NO molecules is formed using a Nijmegen Pulsed Valve that expands 5\% NO seeded in Krypton into vacuum. The Stark decelerator is used to produce packets of NO molecules in the $j=1/2$ rotational level of the $v=0$ vibrational ground state of the $X^2\Pi_{1/2}$ electronic ground state, with a narrow spread around a selected mean velocity. Only the $\Lambda$-doublet level with $f$ parity is selected. The NO particles are state-selectively detected after scattering using a recoil-free (1+1') resonance-enhanced-multiphoton ionization scheme in combination with the new ion optics. The distance from the center of the ionization region to the MCP-detector is around 892.5 mm. The ion optics could be replaced with a reference design for performance comparison. Either tightly focussed lasers were used to detect particles in a small ionization volume, or (partially) unfocussed lasers were used resulting in a large ionization volume (approximately 2.5 mm in all dimensions).

Firstly, the well defined packets of NO $j=1/2,f$ exiting the Stark decelerator at different selected velocities were imaged and used to optimize the voltage settings of the VMI-apparatus experimentally in so called ``beamspot'' measurements. The found optimized voltage settings correspond to applied potentials of around 2000, 1935, 1481 and 0 V for cylinders 1, 2, 8 and 16, respectively. The experimentally obtained settings are thus very close to the predicted settings from the simulations. Furthermore, using a previously established calibration procedure \cite{Zastrow2014}, the velocity mapping was found to be perfectly linear, with each image pixel corresponding to a velocity of 1.14 m/s for the camera configuration used here. For the small and large ionization volume, identical results were obtained. For the large ionization volume, a few volt deviation in the applied potentials did not cause a notable increase in blurring. For the smaller ionization volume, even larger voltage deviations are required to cause a notable increase in blurring. This indicates a stable operation of the VMI-setup at minimal velocity blurring, for both the large and small ionization volume.

To further investigate the performance of the new VMI-setup, the inelastic scattering of the Stark decelerated NO molecules (430 m/s) and a conventional beam of pure He (around 1900 m/s) was used as a model system for molecule-atom collisions. This collision process is known to exhibit narrowly spaced oscillatory structures in the angular distribution of the scattered molecules, originating from the diffraction of matter waves during the collisions \cite{Zastrow2014}. The observation of these diffraction oscillations provides a stringent test for the resolution of a crossed beam scattering experiment, and thus also for the resolution of the VMI-apparatus.

\begin{figure}[!htb]
   \centering
    \resizebox{1.0\linewidth}{!}
    {\includegraphics{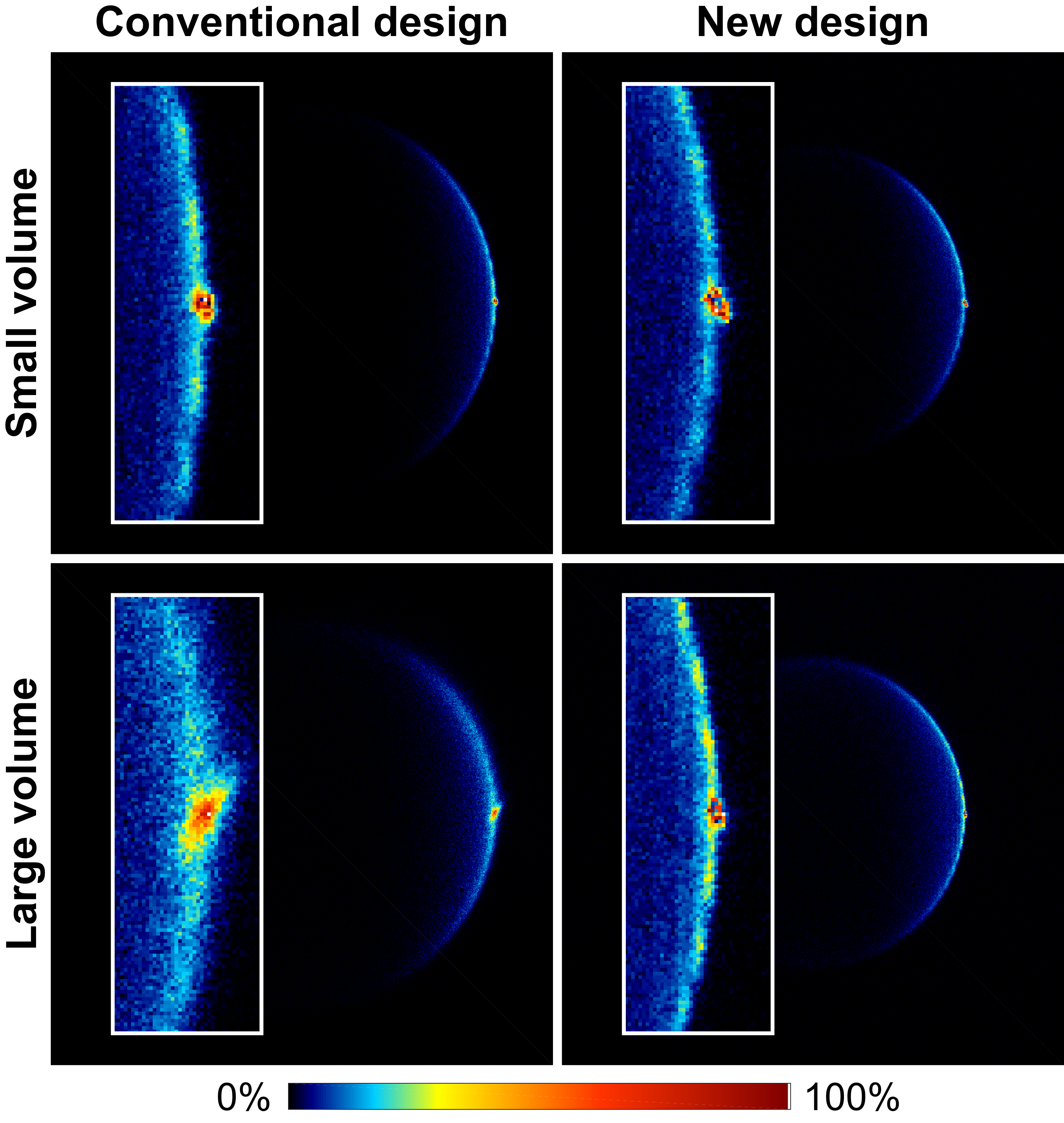}}
    \caption{Velocity distributions of scattered NO molecules as measured with a conventional Eppink-Parker type ion optics design \cite{Eppink1997} and with the new ion optics design, for both small and large ionization volumes. The NO molecules that scattered into the $j=3/2,e$ level were detected after collision with He at an energy of 551 cm$^{-1}$. The insets show an enlarged part of the images near the forward scattering direction. The imaged ring radii are equivalent to about 228 m/s. Each image pixel corresponds to a velocity of 0.92 m/s or 1.14 m/s for the conventional and new design, respectively.}
    \label{fig:New-Conv}
\end{figure}

The images recorded using the new ion optics are shown in figure \ref{fig:New-Conv}, and compared to measurements under identical conditions but using a conventional Eppink-Parker type ion optics design \cite{Eppink1997} with optimized voltage settings. The images are presented such that the relative velocity vector is directed horizontally, with forward scattering angles positioned at the right side of the image. At the forward scattering direction the initial beam gives a contribution to the signal. These measurements were performed at a collision energy of 551 cm$^{-1}$, and with either a small or large ionization volume. To ensure equal comparison of the resolution, the laser powers were reduced for the large ionization volume to give similar ion yields as for the small ionization volume. The image radius for the new apparatus was found to be around 2.4 mm on the MCP-detector plane, in good agreement with the simulated radius of 2.39 mm.

It can be seen that for the small ionization volume sharp images that show clear oscillatory structures in the angular scattering distribution, can be obtained using both the new and conventional ion optics design. For large ionization volumes, however, the image is significantly blurred when using the conventional ion optics, while it remains perfectly sharp for the new design.

For the new ion optics no detrimental effect on the image resolution or beamspot size could be found for the large ionization volume as compared to the small volume. However, the presence of a deteriorative effect is known from both theory and simulations, and is also proven to exist by the experimentally found increased voltage acceptance range for the smaller ionization volume. This indicates that the intrinsic blurring induced by the VMI-apparatus should be well below the measured image resolution, and amounts to at most 1 image pixel or 1.14 m/s. Together with the 228 m/s ring radius, this results in a value of $\sigma<0.7 \%$ for the performed collision experiment, according to the definition of Eq. \ref{blurringformula}, or an intrinsic velocity resolution of $D_V/V'_{NO} \lesssim 0.5 \%$. Thus, the resolution of the scattering image is no longer limited by the blurring of the VMI-apparatus, but by the velocity spreads of the molecular beams and crushing of the Newton spheres, even for large ionization volumes.

\begin{figure}[!htb]
   \centering
    \resizebox{1.0\linewidth}{!}
    {\includegraphics{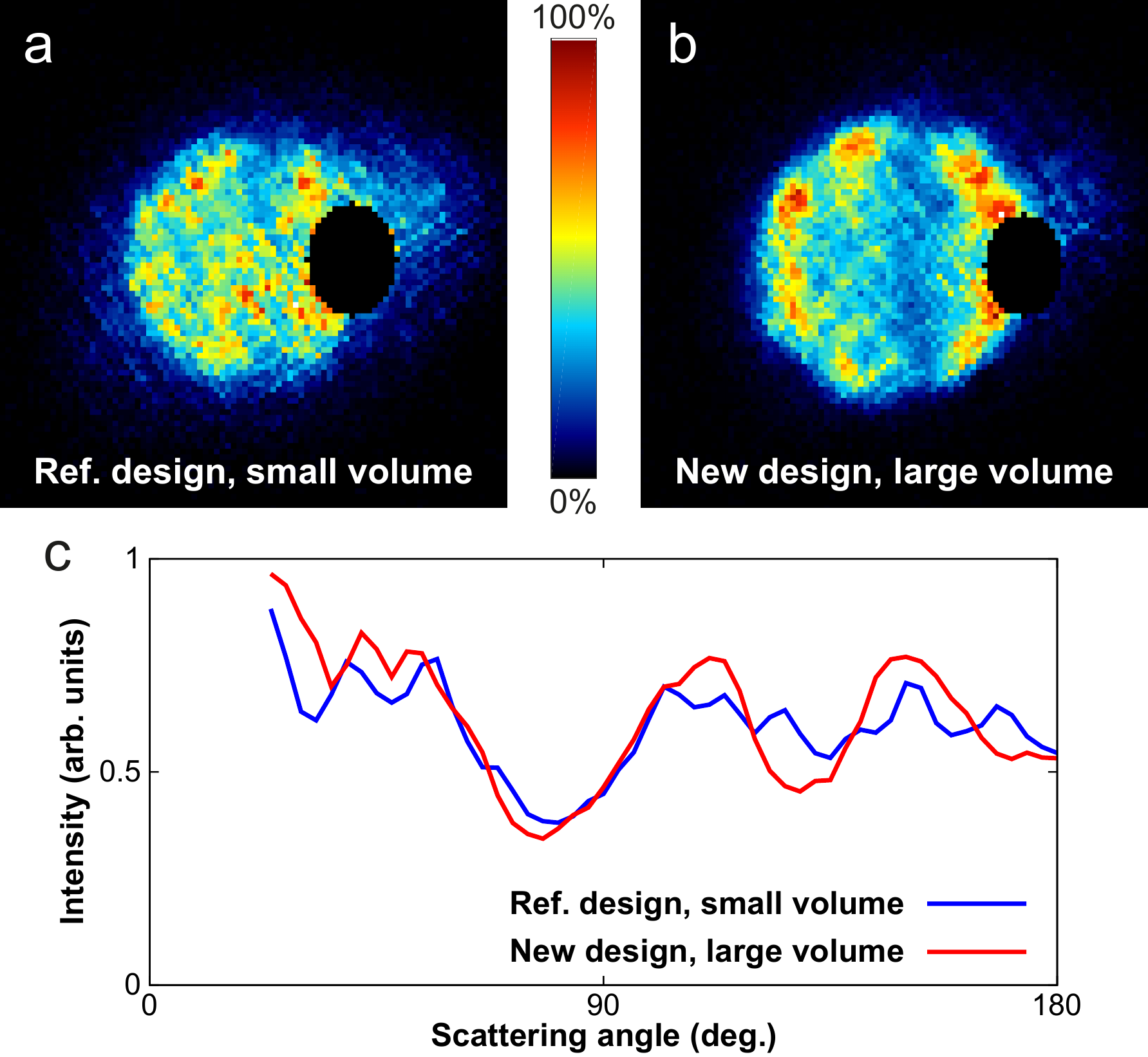}}
    \caption{Velocity distributions of scattered NO molecules as measured with a reference ion optics design adapted from that of Townsend et al. \cite{Townsend2003} when using a small ionization volume (a) and with the new ion optics design when employing a large ionization volume (b). The NO molecules that scattered into the $j=1/2,e$ level were detected after collision with He at an energy of 5.2 cm$^{-1}$. The imaged ring radii are equivalent to about 21 m/s. Each image pixel corresponds to a velocity of 1.08 m/s or 0.94 m/s for the reference and new design, respectively. The angular scattering distributions extracted from the images are depicted in (c).}
    \label{fig:New-Ref}
\end{figure}

To further investigate the mapping accuracy NO-He collisions at a much lower energy of about 5.2 cm$^{-1}$ were studied, using a crossed molecular beam setup optimized for low energy collisions that has been discussed in detail before \cite{Jongh2020}. Here, a Stark decelerated beam of NO molecules (650 m/s) was intersected at an angle of 10$\degree$ with a beam of He (490 m/s) from a cooled Even-Lavie Valve. For this machine the new ion optics design was uniformly scaled with a factor of approximately 1.4. Furthermore, a distance from the ionization region to MCP-detector of around 1133.5 mm was used, and the required potentials were applied with three separate power supplies (i.e. without potentiometer). To illustrate the improvements offered by the new ion optics design for low collision energy experiments, we compare the new design under the least favourable conditions, i.e. using a large ionization volume, with a reference ion optics design adapted from that of Townsend et al.\cite{Townsend2003} under the most favourable conditions, i.e. using a small ionization volume. Apart from the differently sized ionization volume, a corresponding change in laser power to obtain similar signal levels, and a slightly different distance from the ionization region to the MCP-detector (1085 mm for the reference design), otherwise identical experimental conditions were employed. The obtained collision images are depicted in figure \ref{fig:New-Ref}, again presented such that the relative velocity vector is directed horizontally with forward scattering angles positioned at the right side of the image. Small segments of the images are masked around the forward direction since the initial beam gives a contribution to the signal there. The new design gives a much sharper image, even when a much larger ionization volume is used, resulting in a significantly improved contrast between peaks and valleys in the angular scattering distribution extracted from the measured images. For the reference design the use of a large ionization volume resulted in significant image blurring.

The new ion optics design discussed here has recently been employed in several state-of-the-art scattering experiments in our lab, granting the possibility to observe details in the velocity distributions of scattered molecules that had not been resolved before. For example, it allowed resolving narrowly spaced concentric rings, pertaining to product pairs, in the scattering distribution of NO molecules that are excited to high rotational states during collisions with O$_2$ molecules, thus providing experimental data in the largely unexplored field of bimolecular scattering \cite{Gao2018pccp}. More recently, the device has also been employed to image the onset of the resonance regime in low-energy NO-He collisions \cite{Jongh2020}. Here, it allowed measuring the angular scattering distributions at collision energies below 1 K, providing stringent tests for state-of-the-art inelastic scattering calculations in previously unexplored energy regimes. Furthermore, the apparatus has recently been used to capture the first high-resolution images of molecular collisions employing a Zeeman decelerator \cite{Plomp2020}.

\section{Conclusion}
We have presented an improved Velocity Map Imaging ion optics layout, optimized for use in crossed beam scattering experiments employing decelerators. It is particularly designed to provide high-resolution imaging for large detection volumes and over a broad range of collision energies, down to the sub-Kelvin regime. The radially compact setup consists of 16 tube-shaped electrodes, electrically connected via high precision resistors that require only two external power supplies to provide the proper potentials to all electrodes. The experimentally determined operating conditions are in excellent agreement with the simulated ones. The apparatus was found to provide for a negligible intrinsic velocity blurring, even for large ionization volumes of several millimetres in all dimensions. Furthermore, the low electric field strength in the ionization region reduces unwanted mixing of states by the Stark-effect, and provides future prospects for employing direct image slicing techniques. The presented VMI-apparatus design was optimized for the conditions encountered in inelastic scattering experiments employing decelerators but might also find applications in other research areas, for example in the fields of reactive scattering or photodissociation experiments.

\section{Acknowledgments}
This work is part of the research program of the Netherlands Organization for Scientific Research (NWO). S.Y.T.v.d.M. acknowledges support from the European Research Council (ERC) under the European Union's Seventh Framework Program (FP7/2007-2013/ERC Grant Agreement No. 335646 MOLBIL) and from the ERC under the European Union's Horizon 2020 Research and Innovation Program (Grant Agreement No. 817947 FICOMOL). We thank Tim de Jongh and Quan Shuai for their contribution to the measurement and analysis of the low energy collision data. We thank Andr\'e Eppink and David Parker for useful discussions on the VMI technique. We thank Andr\'e van Roij, Niek Janssen and Edwin Sweers for expert technical support.

\bibliographystyle{apsrev}
\bibliography{VMI-biblio.bbl}

\end{document}